\documentclass[structabstract]{aa}  
\usepackage{natbib}
\usepackage{graphicx}
\usepackage{txfonts}
\usepackage{lscape}%

  \def\simge{\mathrel{\raise1.16pt\hbox{$>$}\kern-7.0pt
    \lower3.06pt\hbox{{$\scriptstyle \sim$}}}}           
  \def\simle{\mathrel{\raise1.16pt\hbox{$<$}\kern-7.0pt
    \lower3.06pt\hbox{{$\scriptstyle \sim$}}}}           

\newcommand{\cz}{\ensuremath{C_Z}}
\newcommand{\pv}{\ensuremath{P_V}}
\newcommand{\nv}{\ensuremath{N_V}}

\newcommand{\bz}{\ensuremath{\langle B_z\rangle}}

\newcommand{\nz}{\ensuremath{\langle N_z\rangle}}

\newcommand{\cpn}{CPNs}

\begin{document}
   \title{Magnetic fields in central stars of planetary nebulae?
\thanks{Based on observations collected at the European Organisation for
Astronomical Research in the Southern Hemisphere, Chile, under programme
ID 072.D-0089 (PI=Jordan) and 075.D-0289 (PI=Jordan)}}

   \author{S. Jordan
          \inst{1}
          \and
          S. Bagnulo
          \inst{2}
          \and
          K. Werner
          \inst{3}
          \and 
          S.J. O'Toole
          \inst{4}}
   \institute{Astronomisches Rechen-Institut, Zentrum f\"ur Astronomie der Universit\"at Heidelberg, M\"onchhofstr. 12-14, D-69120 Heidelberg, Germany,
          \email{jordan@ari.uni-heidelberg.de}
         \and
           Armagh Observatory,
           College Hill,
           Armagh BT61 9DG,
           Northern Ireland, U.K.,
           \email{sba@arm.ac.uk}
         \and
         Institute for Astronomy and Astrophysics, Kepler Center for Astro and
          Particle Physics, Eberhard Karls Universit\"at T\"ubingen, Sand 1, D-72076,
T\"ubingen, Germany, \email{werner@astro.uni-tuebingen.de}
         \and  Australian Astronomical Observatory, P.O. Box 296, Epping, NSW
          1710, Australia, \email{otoole@aao.gov.au}
   }

   \date{Received 6 March, 2012; accepted ????, 2012}

 
  \abstract
{
Most of the planetary nebulae (PN) have bipolar or other
non-spherically symmetric shapes. The presence of a magnetic field in
the central star may be the reason for this lack of symmetry, but
observational works published in the literature have so far reported
contradictory results.
}
{
We try to
correlate the presence of a magnetic field with the departures from the
spherical geometry of the envelopes of planetary nebulae.
}
{
We determine the magnetic field from spectropolarimetric observations
of ten central stars of planetary nebulae. The results of the analysis
of the observations of four stars was previously presented and discussed
in the literature, while the observations of six stars, plus additional
measurements for a star previously observed, are presented here
for the first time.
}
{
All our determinations of magnetic field in the central planetary
nebulae are consistent with null results. Our field measurements
have a typical error bar of 150-300\,G. Previous spurious field
detections obtained with FORS were probably due to the use of different
wavelength calibration solutions for frames obtained at different
position angles of the retarder waveplate.
}
{ 
Currently, there is no observational evidence for the presence of
magnetic fields with a strength of the order of hundreds Gauss or
higher in the central stars of planetary nebulae. }
\keywords{stars: post-AGB -- stars: magnetic fields}

   \maketitle
%

\section{Introduction}
There is still no conclusive theory why more than 80\,\%\ of the known planetary
nebulae (PNe) have bipolar and non-spherically symmetric structures
\citep{Zuckerman-Aller:86,Stanghellini-etal:93,CorradiSchwarz:95}. An overview
of possible mechanisms that shape PNe is given by \cite{BalickFrank02}. Several
of these processes imply the presence of magnetic fields which deflect the
outflow of the matter along the magnetic field lines. Recently,
\cite{ThirumalaiHeyl:10} have published model calculations for AGB stars
incorporating both magnetism and stellar winds with dust grains.

Such magnetic fields could be either fossil remnants from previous stages of
stellar evolution, or could be generated by a dynamo at the interface between a
rapidly rotating stellar core and a more slowly rotating envelope.
\cite{Blackman-etal:01} argue that some remnant field anchored in the core will
survive even without a convection zone, although the convective envelope may not
be removed completely.

The idea that magnetic fields are important ingredients shaping PNe has been
supported by the detection of SiO, H$_2$O, and OH MASER emission in
circumstellar envelopes of AGB stars pointing at milligauss fields in these
nebula
\citep{Kemball-Diamond:97,Szymczak-Cohen:97,Miranda-etal:01,Vlemmings-etal:02,Vlemmings-etal:05,Vlemmings-etal:06,Herpin-etal:06,Sabin-etal:07,Kemball-etal:09,
Gomez:09,Vlemmings:11}.  Moreover, using an idea of \cite{Pascoli:85},
\cite{HugginsManley:05} connected the extreme filamentary structures seen in
high-resolution optical images of certain planetary nebulae to magnetic fields
consistent with those measured in the MASER from  the precursor circumstellar
envelopes.

For the first time, and with the help of optical circular
spectropolarimetry carried out with the FORS1 spectrograph of the UT1
(``Antu'') telescope of the Very Large Telescope (VLT) of the European
Southern Observatory (ESO), \cite{Jordan-etal:05} reported on the
detection of magnetic fields in the central stars of the planetary
nebulae NGC\,1360 and LSS\,1362. For the central stars of EGB\,5
and Abell\,36 the existence of a magnetic field was found to be
probable but with less certainty.

\cite{Pascoli:08} pointed out that the magnetic field at the surface of
central stars of planetary nebulae is not necessarily connected to the magnetic
fields in the nebula itself; the latter can be  a fossil component of the
primary field embedded in the AGB star. Nevertheless, the reported detection of magnetic fields in central stars of
planetary nebulae has triggered several additional observational and theoretical
studies on the shaping of planetary nebula, e.g.\,by \cite{Garcia-Diaz-etal:08},
\cite{Tsui:08}, and \cite{PascoliLahoche:10} taking magnetic fields into
account.  On the other hand, \cite{Soker:06} cast strong doubts that magnetic
fields could be the main agent shaping planetary nebulae. He argued  that a
single star cannot supply the energy and angular momentum if the magnetic fields
have a large-scale structure required to shape the outflow from an AGB star.

Recently, the detection of magnetic fields in the central stars of
planetary nebulae (\cpn) was called into question by
\citet{Leone-etal:11}, who re-observed NGC\,1360 and LSS\,1362 with
the FORS2 instrument, and concluded that their effective magnetic
field is null within an uncertainty of $\sim 100$\,G (NGC\,1360), and
$\sim 290$\,G (LSS\,1362). Furthermore, both \citet{Leone-etal:11} and
\citet{Bagnulo-etal:12} re-analysed the observations previously
obtained with FORS1 by \citet{Jordan-etal:05}, and could not confirm
the original detection. 

The conclusion based on the observations of four \cpn\ reduced by two
independent groups is that there is no observational evidence for
magnetic fields in the \cpn. The aim of this work is to enlarge the
sample of \cpn\ checked for magnetic field, in order to estimate the
occurrence of the magnetic field in \cpn\ and, if magnetic field is
detected, whether its presence correlates with the asymmetry of their
envelope. To achieve this goal, two of the teams that had presented
(discordant) results on previous FORS measurements have joint their
efforts to present here a more complete survey of magnetic fields of
\cpn\ obtained during a three-night observing run with FORS1 carried
out in 2005.

\begin{table*}
\caption{\label{Tab_Observations}
Fundamental stellar parameters and FORS1 magnetic field measurements for ten central
stars of planetary nebulae and for a (presumably non magnetic) B9 star observed for comparison.
}
\begin{center}
\begin{tabular}{llrrrrrr@{$\,\pm\,$}lc}
\hline
\hline
\multicolumn{2}{l}{} &     
$T_{\rm eff}$ &              
            &              
Exp. time   &              
Peak SNR    &              
            &              
\multicolumn{2}{c}{\bz} &     
                     \\    
\multicolumn{2}{l}{CPN name/alias} &       
\multicolumn{1}{c}{/K}        &                    
\multicolumn{1}{c}{$\log g$}      &                    
\multicolumn{1}{c}{/s}        &                    
\multicolumn{1}{c}{/\AA$^{-1}$} &                    
\multicolumn{1}{c}{MJD}      &   
\multicolumn{2}{c}{/G} &   
 \multicolumn{1}{c}{remark}                            \\  
\hline
\\
{\bf Programme 072.D-0089:}\\
           &                      &       &     &      &      &           &\multicolumn{2}{c}{} & \\
CD-26\,1389&NGC\,1360             & 97\,000 & 5.3 & 1248 & 1440 & 52946.291 &$  207 $& 325  &1,5 \\
           &                      &       &     & 1248 & 1552 & 52988.235 &$  336 $& 283  &1 \\
           &                      &       &     & 1248 & 1340 & 52989.060 &$  358 $& 361  &1 \\
           &                      &       &     & 1248 & 1400 & 52990.081 &$   72 $& 320  &1 \\
EGB5       &PN G211.9$+$22.6      & 34\,060 & 5.85 & 1986 &  552 & 52988.347 &$ -155 $& 780  &1,6 \\
LSS\,1362   &PN G273.6$+$06.1      & 114\,000 & 5.7 & 1986 &  930 & 52989.309 &$   62 $& 406  &1,5 \\
Abell\,36& ESO 577-24           & 113\,000 & 5.6 & 1500 & 1290 & 53031.287 &$  977 $& 445  &1,5 \\[2mm] 
\hline
\\
{\bf Programme 075.D-0289:}\\
           &                      &       &     &      &      &           &\multicolumn{2}{c}{} & \\
HD\,107969 & NGC\,4361            & 82\,000 & 5.5 & 6000 & 1308 & 53525.093 &$ -348 $& 400  &8\\
           &                      &       &     & 1199 & 1845 & 53526.072 &$   93 $& 295  &\\
           &                      &       &     & 8800 & 1600 & 53527.053 &$  507 $& 351  &\\
Abell\,36& ESO 577-24           & 113\,000 & 5.6& 7200 & 3040 & 53525.004 &$  -33 $& 158  &5\\ 
           &                      &       &     & 3600 & 1880 & 53525.972 &$  110 $& 214  &\\
           &                      &       &     & 3600 & 1880 & 53526.973 &$  115 $& 222  &\\
LSE\,125   &PN\,G335.5$+$12.4     & 85\,000 & 5.1 &10000 & 2950 & 53525.236 &$ -129 $& 116  &9\\
           &                      &       &     & 6000 & 1870 & 53526.208 &$  135 $& 152  &\\
           &                      &       &     & 3600 & 1260 & 53527.266 &$  449 $& 239  &\\
Hen\,2-194 & PN\,H 2-1            & 33\,000 & 3.35 & 3400 &  995 & 53525.325 &$    197 $& 523  & 2,8\\
           &                      &       &     & 6000 & 1250 & 53527.202  &$    -409 $& 544  & 2\\
HD\,154072 & IC\,4637             & 50\,000 & 4.05 & 7200 & 1485 & 53526.295 &$   79 $& 185  &8\\
           &                      &       &     & 4800 & 1110 & 53527.323 &$  -93 $& 216  &\\
HD\,161044 & IC\,1266             & 34\,700  & 3.3 & 6600 & 2820 & 53525.396 &$  -74 $& 123  & 3,10\\ 
           &                      &       &     & 1200 & 1030 & 53526.442 &$ -309 $& 410  & 3 \\
           &                      &       &     & 1200 & 1045 & 53527.425 &$  345 $& 410  & 3 \\
WD\,2226-210&NGC\,7293            & 105\,000 & 7.0 & 6800 & 1065 & 53526.387 &$ 1865 $&1097  &7 \\
           &                      &       &     & 4200 &  920 & 53527.386 &$-1277 $&1269  &\\[2mm]
HD\,160917 &CD-45 11850    &  \multicolumn{1}{c}{--} & \multicolumn{1}{c}{--} &  300 & 3431 & 53527.450 &$  110 $&  68  &4\\
\hline
\end{tabular}
\end{center}

\noindent
1. Observations already published by \citet{Jordan-etal:05}\\
2. The field is estimated from absorption lines only. \\
3. Many spectral lines are in emssion, and the field is estimated from absorption lines
only.\\
4. Non magnetic B9V star observed for comparison.\\
$T_{\rm eff}$ and $\log g$ from: \ \ 5.\cite{Traulsen-etal:05}.\\
\hbox{\ \ \ \ \ \ \ \ \ \ \ \ \ \ \ \ \ \ \ \ \ \ \ \ \ \ \ \ \ \ \ \ \ \ }6. \cite{Lisker-etal:05}.\\
\hbox{\ \ \ \ \ \ \ \ \ \ \ \ \ \ \ \ \ \ \ \ \ \ \ \ \ \ \ \ \ \ \ \ \ \ }7. \cite{Napiwotzki:99}.\\
\hbox{\ \ \ \ \ \ \ \ \ \ \ \ \ \ \ \ \ \ \ \ \ \ \ \ \ \ \ \ \ \ \ \ \ \ }8. \cite{Mendez-etal:92}.\\
\hbox{\ \ \ \ \ \ \ \ \ \ \ \ \ \ \ \ \ \ \ \ \ \ \ \ \ \ \ \ \ \ \ \ }9. \cite{Mendez-etal:88}.\\
\hbox{\ \ \ \ \ \ \ \ \ \ \ \ \ \ \ \ \ \ \ \ \ \ \ \ \ \ \ \ \ \ \ \ }10. \cite{Pottasch-etal:11}.\\

\end{table*}

\section{Observations, data reduction, and magnetic field determinations}

All spectropolarimetric data reported in this paper were taken with
the FORS1 instrument \citep{Appenzeller-etal:98} of the ESO Very Large
Telescope. The polarimetric optics of FORS1, now moved to the twin
instrument FORS2, are based on the principle described by
\cite{Appenzeller:67}. FORS2 now is one of the few optical
spectropolarimeters available for the study of stellar magnetism. Due
to the large aperture of the telescope (8\,m) FORS2 is best suited for
the study of faint stars like white dwarfs
\citep{AznarCuadrado-etal:04,Jordan-etal:07}, sub\-dwarfs
\citep{Jordan-etal:05}, and \cpn\
\citep{Jordan-etal:05}.

The FORS1 dataset previously analysed by \citet{Jordan-etal:05},
\citet{Leone-etal:11}, and \citet{Bagnulo-etal:12} consists of
observations of \object{NGC\,1360}=\object{CD--26\,1340},
\object{EGB\,5}=\object{PN\,G211.9+22.6},
\object{LSS\,1362}=\object{PN\,G273.6+06.1}, and
\object{Abell\,36}=\object{ESO\,577-24}. These observations were
obtained by \cite{Jordan-etal:05}, in service mode between November
2, 2003, and January 27, 2004, using grism 600B, and a 0.8\arcsec\
slit width, for a spectral resolution of about 1000.

Three additional nights of telescope time were obtained in visitor
mode between June 4, 2005, and June 6, 2005, at the UT2 (``Kueyen'')
of the ESO VLT. The instrument setup included again grism 600\,B, but
a slit width of 0.7\arcsec, for a spectral resolution of about 1200.

The criterion for the selection of objects were the optimum visibility during the
observation nights and the brightness of the stars in order to reach an optimum signal-to-noise ratio. 
Moreover, we made sure that at least two of the stars were in the  center of a nebula with
almost spherical shape (\object{LSE\,125} and  \object{Hen\,2-194})
so that in case of positive detections the magnetic field can be  correlated with the 
topology of the nebulae.

During the visitor run, six new \cpn\
were observed two or three times: \object{HD\,107969}=\object{NGC\,4361},
\object{LSE\,125}=\object{PN\,G335.5+12.4},
\object{Hen\,2-194}=\object{ESO\,392-2},
\object{HD\,154072}=\object{IC\,4637},
\object{HD\,161044}=\object{IC\,1266}, and
\object{WD\,2226-210}=\object{NGC\,7293}.
\object{Abell\,36}=\object{ESO\,577-24}, already checked for magnetic
field in the previous service observing run, was re-observed three
times. Additionally, we took one spectropolarimetric dataset for
\object{HD\,160917}, a non-magnetic B9V comparison star.

\begin{figure}[t]
\includegraphics[width=0.5\textwidth,angle=0]{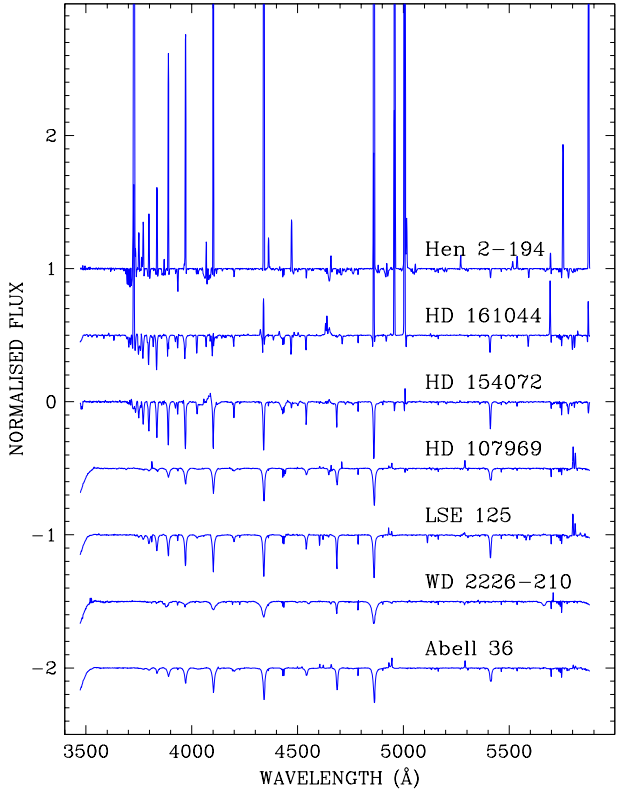}\\
\caption{\label{Fig_spectra} 
Normalised spectra of all central stars of planetary nebulae observed in the  075.D-0289 campaign ordered by increasing effective temperature
from top to the bottom. The 072.D-0089 are shown in Fig.\,1 of \cite{Jordan-etal:05}.
}
\end{figure}

Figure\,\ref{Fig_spectra} shows the summed up high-quality  spectra for all
\cpn. Only \object{Hen\,2-194} and \object{HD\,161044} were subject to 
significant nebula emission.

We analysed all this old and new observational material
by adopting a method described in detail by \citet{Bagnulo-etal:12}.
Spectra are calibrated and extracted using the ESO FORS pipeline
\citep{Izzo-etal:10}, then combined using the difference
method to obtain both the reduced Stokes $V$ profiles (\pv) and the
null profiles (\nv), as described by \citet{Bagnulo-etal:12}.  The mean
longitudinal magnetic field \bz\ was then calculated by using a
least-square technique based on the relationship
\begin{equation}
\pv(\lambda)= - g_\mathrm{eff} \ \cz \ \lambda^{2} \
                \frac{1}{I(\lambda)} \
                \frac{\mathrm{d}I(\lambda)}{\mathrm{d}\lambda} \
                \bz\;,
\label{Eq_Bz}
\end{equation}
where \pv\ is the reduced Stokes $V$ profile, $I(\lambda)$ is the Stokes $I$
profile of a spectral line, $g_\mathrm{eff}$ is the effective Land\'{e} factor, and
\begin{equation}
\cz \simeq 4.67 \times 10^{-13}\,\AA^{-1}\ \mathrm{G}^{-1} \; .
\end{equation}
For more details, see, e.g., \citet{Bagnulo-etal:12}.
Compared with \citet{Bagnulo-etal:12}, in this work we have implemented
a sigma-clipping algorithm in the determination of the magnetic field from
the correlation diagram of circular polarisation against local flux
derivative. We have also calculated the null profiles \nv\ and compared their
oscillation about zero with the \pv\ error bars, and also measured the
null field \nz, i.e., the magnetic field obtained by applying Eq.~(\ref{Eq_Bz})
to the \nv\ profile instead of \pv. The statistical significance of the
null field values is extensively discussed in \citet{Bagnulo-etal:12}. Here we
report that all null field values were found consistent with zero within error bars.
We also note that the targets of this survey are relatively faint, and that most of 
the observations discussed here are not characterised by a ultra-high signal-to-noise
ratio. The main contributor to the error bars is thus photon-noise and background 
subtraction (since our targets are embedded in a circumstellar envelope).

The original data reduction of the observations obtained within
Programme ID 072.D-0089 by \cite{Jordan-etal:05} was based on two
distinct wavelength calibration solutions for the frames obtained at
the two different position angles of the retarder plate adopted for
the science observations: science frames obtained with the retarder
waveplate at position angle $-45\degr$ were calibrated with
calibration frames obtained with the retarder waveplate at $-45\degr$,
and science frames obtained with the retarder waveplate at $+45\degr$
were calibrated with calibration frames obtained at
$+45\degr$. However, \citet{Bagnulo-etal:06} and
\citet{Bagnulo-etal:09} shown that this method prevents wavelength
calibration errors to cancel out, and may lead to spurious
detections. While this problem did not occurr in various tests
\citep[e.g.][]{Bagnulo-etal:02}, in the case of for instance the
075.D-0289 data it would lead to spurious magnetic field
measurements of the order of $\sim 9$\,kG.

In their analysis of the data obtained in 072.D-0089,
\cite{Jordan-etal:05} did compare the Stokes profiles obtained using a
common wavelength calibration with those obtained using two distinct
solutions, and concluded that both methods would produce similar
Stokes profiles. However, at that time they did not notice that,
although the profiles looked similar, field measurements were
different: after the analysis of the profiles obtained using two
wavelength calibration frames, a field was firmly detected, while
using a unique wavelength calibration frame, field detections would
disappear. We conclude that for that reason, the relatively high
magnetic fields reported by \cite{Jordan-etal:05} are spurious.

A similar conclusion was drawn by \citet{Bagnulo-etal:12} and
\citet{Landstreet-etal:12} for the measurements of magnetic fields in
subdwarfs by \citet{OToole-etal:05}: the kG magnetic fields are also
disappearing when using a single wavelength calibration frame for the
entire science dataset. We note instead that the detections of weak
magnetic fields in white dwarfs by \citet{AznarCuadrado-etal:04} were
basically confirmed by \citet{Bagnulo-etal:12}.

\section{Results}\label{Sect:Results}

\begin{figure}[t]
\includegraphics[width=0.39\textwidth,angle=270]{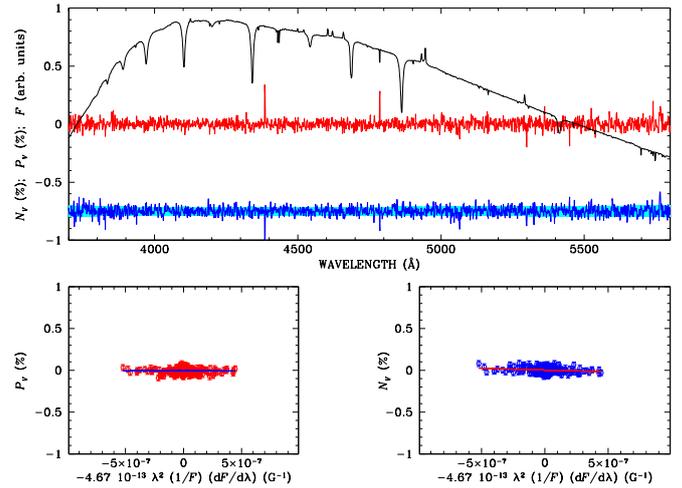}\\
\caption{\label{Fig_Abell36} 
The observations of Abell\,36 were obtained with FORS1 on 2005-06-03.  The
top panel shows the observed flux $F$ (black solid line, in arbitrary
units, and not corrected for the instrument response), the
$\pv = V/I$ profile (red solid line centred about 0), and the null profile
\nv\ (blue solid line, offset by $-0.75$\,\% for display purpose). The null
profile is expected to be centred about zero and scattered according
to a Gaussian with $\sigma$ given by the \pv\ error bars. The
\pv\ error bars are represented with light blue bars centred about
$-0.75$\,\%.  The slope of the interpolating lines in the bottom panels
gives the mean longitudinal field from \pv\ (left bottom panel) and
from the null profile (right bottom panel) both calculated using the H
Balmer and metal lines. The corresponding \bz\ and \nz\ values are $-33
\pm 158$\,G and $-420 \pm 161$\,G, respectively.
}
\end{figure}

Table\,\ref{Tab_Observations} lists our field determinations from all
new FORS observations. We have also included new field determinations
from the observations carried out in service mode by
\citet{Jordan-etal:05}.  Note that the field estimates for the
observations obtained in 072.D-0089 slightly differ from those
published by \citet{Bagnulo-etal:12} because of the implementation of
the sigma-clipping algorithm, and also a slightly different choice of
the spectral points considered for field
determination. Figure\,\ref{Fig_Abell36} shows an example of
reduced data.

\section{Discussion}
Our analysis of ten \cpn\ does not show significant evidence for the
existence of longitudinal magnetic fields above a few hundred
Gauss. This is in contradiction with the result of
\cite{Jordan-etal:05}, who determined magnetic field strengths of
several kG in their analysis of four \cpn. Our current results are
consistent with the investigation of the central star of
NGC\,1360 by \citet{Leone-etal:11}, who determined 
an upper limit for
the magnetic field of 300\,G \citep[while][reported a longitudinal
magnetic field of up to 3\,kG]{Jordan-etal:05}. For the central star of
LSS\,1362  \citet{Leone-etal:11} obtained an upper limit of 600\,G.
We conclude that a non-optimal wavelength calibration method has led
\citet{Jordan-etal:05} to a spurious detection of magnetic fields.

Our re-reduction of the 072.D-0089 data leads to a weighted mean
magnetic field (see Table\,\ref{Tab_Observations}) for NGC\,1360 of
$244\pm162$\,G, assuming that a possible (and weak) magnetic field
appears constant with time.

With our sample of ten stars (Abell\,36 has been observed in both
observation campaigns) we conclude that there is no confirmed case for
a magnetic field in the central star of a planetary nebuale at a kG
level. Magnetic fields of the order to 100-300\,G, however, cannot be
discarded. Indirect evidence for the existence of mG fields in
proto-planetary nebulae could still support an influence of magnetic
fields on the shape of PN. However, these fields need no be connected
to a field of the central star \citep{Pascoli:08}.

\acknowledgements{We thank the staff of the ESO VLT for carrying out
  the service observations of programme 072.D-0089 and supporting our
  visitor mode observations of programme 075.D-0289 (PI=S. Jordan for
  both programmes).}

\bibliographystyle{aa}
\bibliography{cpn.bbl}

\end{document}